\documentstyle{article}

\begin{document}
\title {Coherent Emission from Magnetars}
\author
{\\ David Eichler,$^1$ Michael Gedalin,$^2$  and Yuri
Lyubarsky$^3$
 \\ Department of  Physics \\
Ben-Gurion
University \\ Beer-Sheva, 84105, Israel \\
1. eichler@bgumail.bgu.ac.il \\ 2.
gedalin@bgumail.bgu.ac.il\\ 3. lyub@bgumail.bgu.ac.il}
\date{Accepted xx. Received xx}

\maketitle
\begin{abstract}
     It is proposed that magnetospheric currents above the surfaces of
magnetars radiate coherent emission in analogy to pulsars. Scaling
the magnetospheric parameters suggests that the coherent emission
from magnetars  would emerge in the infra-red or optical.

\end{abstract}

Pulsar radio emission and molecular masers are two examples of
naturally occurring coherent emission. But there is no accepted
analogue in the optical to our knowledge.
 A brief optical flash  was detected from gamma ray burst GRB 990123, but
it  can be explained as being incoherent if the bulk Lorentz
factor of the expanding fireball thought to generate is above 50
(Akerlof et al, 1999).

 In this letter we consider the
possibility of {\it coherent} optical and IR emission in certain
astrophysical situations, such as magnetars. The basis for such a
hypothesis is quite simple:
 Magnetars, it has been proposed (Duncan and Thompson, 1992, Thompson
and Duncan, 1995, 1996), have twisted
magnetic loops in their magnetospheres, and, most of the time, the
thermal scale height of their atmospheres is too low to populate
the magnetosphere with thermal plasma. On the other hand,
magnetospheric currents can easily be drawn out of the surface of
the star from at least one of the footpoints. This current can be
estimated.  The rate of dissipation dB/dt of the magnetic field
can also be estimated; it is presumably the value of dB/dt that
yields a sufficient potential
 drop across the length
of the loop to create enough plasma to   short out any larger
potential
 drop. The
density of plasma so estimated is many orders of magnitude larger
than that in pulsar magnetospheres. If pulsars can radiate
coherently in the radio, this frequency being ultimately
determined by the plasma frequency in the pulsar magnetosphere,
then similar processes could occur in magnetar magnetospheres with
the  plasma frequency scaled up appropriately.

\section{Basic Numbers}
    Consider an arched magnetic flux tube,
    similar in shape to a solar prominence,
 twisted to a pitch angle of   unity over a
 scale of
$10^6R_6^*$ cm, where the longitudinal field along the flux tube
is defined to be $10^{15}B^*_{15}$ G. We assume that the arch is
stable - perhaps a remnant of what was once a more twisted, less
stable one that flared. The maximum pitch of twisting,  which
presumably occurs at the top of the loop where $B^*_{15}\ll 1$, is
then about unity there and less at lower altitudes. This implies
an energy in the twisted field of order $10^{46-47}B^{*2}_{15}
R_6^{*3}$ergs, a current density of

\begin{equation}
j = \frac{ec}{4\pi} \times [4 \times 10^{18} (B^*_{15}/R_6^*)
cm^{-3}],
\end{equation}
a total current of $I = 3 \times 10^{40}eB^*_{15}R_6^*$
electrons/s, and a rate of energy dissipation that is at least
 $I\Phi$ where  $\Phi $ is the potential drop across the magnetic arch.

If the current-bearing plasma in the arch is single species, then
the electric field it would create is of order $10^{15}$
Statvolt/cm,  which is impossibly high, so there must be
 quasi-neutrality in the arch.
  If the charge balance is to be
 maintained by ions, then the potential drop across the arch must be at
 least
of order 100 MeV, to raise ions well above the neutron star
surface.  If it is to be maintained by pair creation,
 then similar potential drops must occur so that the pairs can either
 a) curvature radiate
 gamma rays of
order 10 Mev to make more pairs, or b) inverse- Comptonize the
thermal photons from the surface to 10 MeV or so.

That a two-species plasma is needed suggests that there is
counterstreaming between the positive and negative charges. This
gives rise to a broad band two stream instability at a frequencies
less than  $\omega_*\equiv 2\omega_p\gamma^{1/2}$ (for a recent
review, see Lyubarsky, 2002),
  where $\gamma$ is the Lorentz
factor and $\omega_p$ is the plasma frequency of the outflowing
plasma in the laboratory frame. The two stream instability
produces electrostatic waves that are converted by non-linear
processes or field line curvature to electromagnetic waves
(Lyubarsky, 2002). Alternatively,   with nearly the same growth
rate, it can  produce slightly oblique subluminous waves that
convert to superluminal ones via resonance broadening at the
height where the real parts of the frequency are closest to each
other ($\omega=\omega_*$) if at this point the frequency
difference between  the two branches is of the order of the growth
rate (Gedalin, Gruman, and Melrose, 2002). In a hot plasma
subluminous waves cannot propagate at frequencies higher than some
cutoff frequency $\omega_c \geq  \omega_*$, which depends on the
details of the distribution function
 (Gedalin, Gruman, and Melrose, 1998). When the point
$\omega=\omega_c$ is reached, the subluminal wave experiences
efficient refractive conversion into the superluminal mode, which
is the only propagating mode in this region (Gedalin, Gruman, and
Melrose, in preparation). These latter authors emphasize that the
waves can grow non-resonantly in a broadband below $\omega_*$ and
convert via direct, linear conversion to the non-growing but
freely escaping superluminal branch at the height where
$\omega=\omega_*$ (or $\omega=\omega_c$). Thus lower frequencies
escape at higher altitudes and enjoy a longer growth path and,
hence, a higher gain factor.

Regardless of the details of any particular model for the coherent
emission, escaping coherent radiation probably has a minimum
frequency of $\omega_p^{\prime}$ in the frame of the
 outflowing plasma, which gives it  a frequency in the observer
frame of $2\omega_p \gamma^{1/2}$.

        We now attempt to estimate the values of $\omega_p$
        (equivalently, the density)
and $\gamma$ for a magnetic arch in the magnetosphere of a
magnetar. The total density must clearly be at least as high as
the minimum to deliver the required current $c\nabla \times
B/4\pi$ but, as charges of both sign are required to avoid
absurdly high electric fields, it is probably much higher by a
multiplicity factor $\eta$. Estimating this factor is difficult,
and, in standard pulsar theory, estimates range from $\sim1$ to
$10^6$ (e.g. Hibschmann and Arons 2001). Usually, the
characteristic Lorentz factor $\gamma$ is estimated as being that
which will give rise to charges of both signs, which is a
necessary condition for shorting out the strong electric fields
that would otherwise obtain.
    Pairs can be produced via gamma rays interacting with the strong
magnetic field.  The gamma rays  can be produced by curvature
radiation of primary  electrons only if the Lorentz factor of the
latter is of order $10^6$. A more likely mechanism is the resonant
scattering of thermal X-ray photons the emerge from the star's
surface. In order to be resonantly scattered,  thermal X-mode
photons of energy $\epsilon_{\gamma}$ must have the Landau kinetic
energy $([1+ 2B/B_{QED}]^{1/2}-1) m_ec^2$ in the rest frame of the
electron, where $B_{QED}=4.4\times 10^{13}G)$. Because the Landau
energy $E_L$ is  about 3$B_{15}^{1/2}$ MeV  when $B\gg B_{QED}$, a
Lorentz factor of $\gamma \sim E_L/\epsilon \sim 10^{2.5}(10
KeV/\epsilon) B_{15}^{1/2}$ is needed. Far from the magnetar
surface, where $B_{15}\ll 1$, a lower $\gamma$ would be needed for
resonant scattering, and the observed non-thermal X-ray spectrum
can be attributed to more gentle inverse-Compton boosting at high
latitudes (Thompson, in preparation). However, the energy
dissipation in the loop must  be determined at the point of
highest potential drop, which in this scenario  would be at the
footpoints, where $E_L$ is highest.
     Because the ratio $\Lambda$ of photons to electrons in the
arch  is  more than $10^{10}$, $\epsilon$ can be chosen to be
about $ln\Lambda \sim  25$ times $kT_s$, where $T_s$ is the
surface temperature, about 0.4 KeV.  Hence $\epsilon \sim  10$ KeV. Photons
scattered down to the surface from higher altitudes might arrive
at even larger energies in sufficient number to serve as seeds
for pair production.

Equation (1)  predicts a minimum dissipation rate of
$10^{36.5}\gamma_{2.5} B_{15}^*R_6^*$ erg/s (where
$\gamma_{2.5}\equiv \gamma/10^{2.5}$), which, when compared to the
observed rate of persistent emission from magnetars, $\sim
10^{35}L_{35}$ erg/s, even less in the pulsed component,  suggests
that at the highest point on the arch,

\begin{equation}
 B_{15}^*R_6^*  \le 10^{-1.5}/\gamma_{2.5},
\end{equation}
 This estimate is supported by the
energetic consideration that the energy in the twisted field,
which  is of order $10^{46-47}B_{15}^{*2} R_6^{*3}$ ergs, is not
likely to exceed the energy of the giant flare itself $\sim
10^{44-45}$ ergs. The magnetic field at the top of the loop is
probably small compared to the surface field of nearly $10^{15}$
G, i.e $B_{15}\ll 1$. (The alternative way of satisfying  the
above equation, that $R_6\ll 1$, is less attractive given the
non-thermal spectrum of the X-ray emission, which is accounted for
by resonant cyclotron scattering of photons in a  highly twisted
field of order 10$^{12-13}$ Gauss. Rather, the picture is that
loops balloon up to altitudes comparable to or greater than the
magnetar radius, and so cover enough  of the surface to affect the
spectrum)

 By equation (1) and the assumption that the particles move at a constant
velocity c, the plasma frequency
$(4\pi e^2 n/m_e)^{1/2}$, where $m_e$ is the rest mass of the
electron, is
\begin{equation}
\omega_{p} = 1 \times 10^{14} (\eta B_{15}^*/R_6^*)^{
\frac{1}{2}}(R_6^*/R_6)hz
\end{equation}
Because  $BR^2$ has the  constant value $B^*R^{*2}$
 along the flux tube we can rewrite the
above equation as
\begin{equation}
\omega_{p} = 1 \times 10^{14}
 R_6^{*-\frac{1}{2}}(\eta B_{15})^{ \frac{1}{2}}hz.
\end{equation}

We  then estimate that the frequency for coherent emission from
the footpoint of the magnetic arch is
\begin{equation}
\nu_m \sim  2\gamma^{ \frac{1}{2}}\omega_{p}/2\pi  \sim 1 \times
10^{15}(\eta B_{15}\gamma_{2.5})^{\frac{1}{2}}{R_6^*}^{
\frac{1}{2}}hz
\end{equation}
Near the surface, where $B_{15}\sim 1$ , this suggest emission in
the near IR, optical, or even UV for high enough $\eta$.

A similar estimate for pulsars, using the Goldreich-Julian density
$(\Omega B/2\pi c)$  would
yield about
\begin{equation}
\nu_p \sim 3\times 10^{10} [\gamma_{2.5} B_{12} \Omega_0
\eta]^{\frac{1}{2}} hz,
\end{equation}
where $\Omega_0$ is the angular frequency of the pulsar's spin in
seconds. This decreases with height from the pulsar surface, and
pulsar emission at hundreds of Mhz is presumed to arrive from
about 10 stellar radii, where $B_{12}\sim 10^{-3}$. Emission at
higher frequencies, however, is typically seen, and occasionally
extends out to about 30 Ghz. The angle of the emission cone is
typically larger at lower frequencies, and this suggests that many
or most
 spectral breaks below   $\nu_p$ are due to the line of sight missing the
pulse at the highest frequencies.

Similarly, a twisted magnetic arch that protrudes from a
magnetar surface could
emit over a broad band, depending on the exact altitude of the
emission.

\section{Discussion}

The above estimate suggests that coherent emission can be emitted
from the magnetospheres of magnetars in the near infra-red or
optical. As the emission is likely to be beamed, in analogy with
pulsars, we cannot predict that any given magnetar would be a
coherent IR or optical  source, but there are enough in the Galaxy
- assuming this includes anomalous X-ray pulsars - that one or
several might be. Because many are in the plane of the Galaxy, the
near IR might be a good frequency to search for such emission.

 Magnetars should be distinguished (at least within the framework of
the present hypothesis) from pulsars in that their current carries
a much larger fraction of the total energy budget than the latter.
This view is supported by the fact that the pulsed component of
the persistent X-ray flux is typically at least 10 percent or so
of the total. Considering that the emission is non-thermal and
probably inverse-Comptonized in the magnetosphere (Thompson 2002),
we conclude that most of the energy budget passes through the
magnetospheric currents.  Thus, the important question of
efficiency needs to be defined carefully. The coherent emission of
pulsars is only a small fraction of the spin-down power, but it
can be a much higher fraction of the power in polar currents, as
the latter is itself only a small fraction of the total. By the
same token, a considerable fraction of the long term magnetic
energy dissipation  in magnetars could end up as coherent
electromagnetic emission; 1 to 10 percent is not unreasonable. A
magnetar at a distance of 5 to 10 kpc could be detectable at 2.2
microns with the next generation technology even if $10^{33}$
erg/s ( $\sim 10^{-2}$  of its persistent X-ray flux).
    Such emission would almost certainly have the period of the
magnetar, and would probably be polarized. In analogy to pulsars,
where the direction of polarization can swing with pulse phase,
the time-integrated polarization would probably be less than at
any instant, but it could nevertheless be non-zero.

Similarly, $10^{33}$erg/s (or less) in coherent optical emission
from SGR 0526-66 in the LMC would be  about 25th magnitude (or
fainter) and might be detectable.

We have considered the possibility that the recently reported
optical emission (Hulleman, van Kerkwijk, and Kulkarni 2000)from
the anomalous X-ray pulsar (AXP) 4U0142+61 is coherent. Since this
paper was first drafted, the optical emission has been reported
(Kern and Martin, 2002) to have the periodicity of the AXP. Data
on polarization or upper limits have not to our knowledge been
reported yet. Although the optical radiation is arguably
non-thermal, general energetic and thermodynamical considerations
still allow it to be incoherent (e.g. Eichler and Beskin, 2000).
On the other hand, the particular plasma mechanisms that give
coherent radiation at low frequencies may  turn out to explain the
data well. The high pulsed fraction of the optical emission -
0.27, considerably higher than that of the soft X-ray  emission -
is naturally explained by a pulsar-type emission mechanism.
Synchrotron or cyclotron emission would either have to be at much
shorter wavelengths  or quite far from the magnetar surface.

\section{Acknowledgements}

 We thank C. Thompson for helpful
discussions. Support from the Arnow Chair of Physics and an Adler
Fellowship awarded by the Israel Academy of Sciences is
acknowledged with gratitude.

\end{document}